\documentclass[12pt,preprint]{aastex}

\usepackage{natbib}
\usepackage{morefloats}
\usepackage{graphicx}

\slugcomment{Edit 12 May 2009}

\shorttitle{Magnetic-Field Morphology in W51 e2/e8}
\shortauthors{Tang et al.}

\begin{document}


\title{Evolution of Magnetic Fields in High Mass Star Formation:
Linking field geometry and collapse for the W51 e2/e8 cores}

\author{Ya-Wen Tang}
\affil{Department of Physics, National Taiwan
University, No. 1, Sec. 4, Roosevelt Road, Taipei 10617, Taiwan \&\\
Academia Sinica Institute of Astronomy and Astrophysics, P. B. Box
23-141, Taipei 10617,Taiwan}

\author{Paul T. P. Ho}

\affil{Academia Sinica Institute of Astronomy and Astrophysics, P.
B. Box 23-141, Taipei 10617,Taiwan \&\\
Harvard-Smithsonian Center for Astrophysics,60 Garden Street
Cambridge, MA 02138, U.S.A.}

\author{Patrick M. Koch}

\affil{Academia Sinica Institute of Astronomy and Astrophysics, P.
B. Box 23-141, Taipei 10617,Taiwan}

\author{Josep M. Girart}
\affil{Institut de Ci\`{e}ncies de l'Espai (CSIC-IEEC), Campus
UAB, Facultat de Ciencies, Torre C5 - parell 2, 08193 Bellaterra,
Catalunya, Spain}

\author{Shih-Ping Lai}
\affil{Academia Sinica Institute of Astronomy and Astrophysics, P.
B. Box 23-141, Taipei 10617,Taiwan \&\\
Institute of Astronomy and Department of Physics, National Tsing
Hua University, 101, Section 2, Kuang Fu Road, Hsinchu, Taiwan
300, R. O. C.}

\author{Ramprasad Rao}
\affil{Submillimeter Array, Academia Sinica Institute of Astronomy
and Astrophysics, 645 N. Aohoku P1, HI 9672, USA}


\begin{abstract}
We report our observational results of 870 $\mu$m continuum
emission and its linear polarization in the massive star formation
site W51 e2/e8. Inferred from the linear polarization maps, the
magnetic field in the plane of sky (B$_{\bot}$) is traced with an
angular resolution of 0$\farcs$7 with the Submillimeter Array
(SMA). Whereas previous BIMA observations with an angular
resolution of 3$\arcsec$ (0.1 pc) showed a uniform B field, our
revealed B$_{\bot}$ morphology is hourglass-like in the collapsing
core near the Ultracompact H II region e2 and also possibly in e8.
The decrease in polarization near the continuum peak seen at lower
angular resolution is apparently due to the more complex
structures at smaller scales. In e2, the pinched direction of the
hourglass-like B field morphology is parallel to the plane of the
ionized accretion flow traced by H53$\alpha$, suggesting that the
massive stars are formed via processes similar to the low mass
stars, i.e. accretion through a disk, except that the mass
involved is much larger. Furthermore, our finding that the
resolved collapsing cores in e2 and e8 lie within one subcritical
0.5 pc envelope supports the scenario of \textit{magnetic
fragmentation} via ambipolar diffusion. We therefore suggest that
magnetic fields control the dynamical evolution of the envelope
and cores in W51 e2 and e8.

\end{abstract}



\section{Introduction}
The magnetic (B) field has been suggested to play an important
role in the star formation process. While the B field flux density
is eventually redistributed via ambipolar diffusion
\citep{mestel56, mouschovias78}, the collapse itself is slowed
sufficiently to explain the low star formation rate observed in
molecular clouds. Alternative support via turbulence
\citep[cf.][]{maclow04} seems less important on parsec (pc) scales
since the B fields in the plane of sky (B$_{\bot}$) are often
observed to be organized and uniform across the cloud, such as in
M17 \citep{dotson96}, OMC-1 \citep{schleuning98} and DR21 MAIN
\citep{kirby09}. One key question is at which sizescale will the
magnetic support be overcome by gravity. The morphology of the B
field at that point may reveal the details of the contraction
process such as geometry and timescale. The Submillimeter Array
(SMA) can be used to address this question by resolving the
B$_{\bot}$ structures via dust polarization studies with high
angular resolutions at typically a few arcseconds.

The B field is traced by the dust continuum emission. The dust
grains are most likely not spherical in shape, but somewhat
elongated. They are thought to be aligned with their minor axes
parallel to the B field in most of the cases \citep{lazarian07}.
Among different alignment mechanisms, radiation torques seem to be
a promising mechanism to align the dust grains with the B field
\citep{draine96,lazarian072}. Due to the differences in emissivity
perpendicular and parallel to the direction of alignment, the
observed thermal dust emission will be linearly polarized. The
direction of the linear polarization is therefore perpendicular to
the B field. With the SMA, we are able to detect the polarized
component of the thermal dust emission at sub-millimeter (sub-mm)
wavelengths in order to trace the B field within the dense cores,
where stars are formed. Compared to the polarization studies via
absorption and scattering of stellar light in the optical or near
infrared \citep{goodman95}, sub-mm polarization, being derived
directly from dust emission, does not suffer from a limited range
in grain size and the possible contamination from the more diffuse
emission and absorption along the line of sight.

In this paper, we present SMA observational results with an
angular resolution of 0$\farcs$7 (0.02 pc) of the massive star
forming site W51 e2 and e8 in W51 MAIN. The dust continuum at a
wavelength of 870 $\mu$m, and the B$_{\bot}$ field inferred from
its linearly polarized component are presented. The W51 MAIN is on
the eastern edge of W51. It is at a distance of 7.0$\pm$1.5 kpc
\citep{genzel81} or 6.1$\pm$1.3 kpc \citep{imai02}. Here, we adopt
a distance of 7 kpc. There is a group of Ultracompact H II (UCHII)
regions in W51 MAIN, and many H$_{2}$O, OH and NH$_{3}$ maser
spots have been identified \citep{genzel81,gaume87,pratap91} to be
associated with the e2 and e8 regions. The terminology of the
structures discussed in this paper is shown in the schematics in
Figure 1. The radio continuum sources e2, e4, e8, e1 and e3 are
UCHII regions \citep{gaume93,zhang97}, and their locations are
labelled in Figure 2. Hereafter, \textit{e2}, \textit{e4},
\textit{e8}, \textit{e1} and \textit{e3} (when in \textit{italic})
refer to their corresponding UCHII regions. The infall signatures
toward the e2 and e8 regions (i.e., e2 and e8 collapsing cores)
have been detected clearly in NH$_{3}$ \citep{ho96,zhang97} and in
CS \citep{zhang98}, indicating that they are in an early
evolutionary stage. Furthermore, the total luminosity of the W51
MAIN is 2$\times$10$^{6}$ L$_{\sun}$ \citep{jaffe87}, indicating
that it is a massive star forming site.

The polarized dust emissions associated with the envelope of the
W51 e2 and e8 regions have been previously observed at 1.3 mm and
850 $\mu$m. The B$_{\bot}$ field structure varies with different
size scales. Chrysostomou et al. (2002) has shown that the
morphology of the field on the very large scale observed with
SCUBA with an angular resolution of $\sim$ 10$\arcsec$ (0.5 pc)
appears more complex, possibly because of projection effects from
several clouds along the line of sight. With an angular resolution
of $\sim$3$\arcsec$ (0.1 pc) with the BIMA, Lai et al. (2001)
found that the position angles (P.A.s) of the polarization vectors
vary smoothly across the e2 and e8 cores (Figure 2(a)), suggesting
that the B field dominates over the turbulent motions in the
envelope. At which scales will the B field lose its dominance over
turbulence and gravity?



\section{Observation}
The observations were carried out on 2008 July 13 using the SMA
\citep{ho04}\footnote{The Submillimeter Array is a joint project
between the Smithsonian Astrophysical Observatory and the Academia
Sinica Institute of Astronomy and Astrophysics and is funded by
the Smithsonian Institution and the Academia Sinica.} in the
extended configuration, with seven of the eight antennas
available. The projected lengths of baselines ranged from 30 to
262 k$\lambda$. The largest size scale which could be sampled in
this observation was $\sim$8$\arcsec$ (0.3 pc). The local
oscillator frequency was tuned to 341.482 GHz. With the 2 GHz
bandwidth in each sideband, we were able to cover the frequency
ranging from 345.5 to 347.5 GHz and from 335.5 to 337.5 GHz in the
upper and lower sidebands, respectively. The phase center is near
\textit{e2} at Right Ascension (J2000) =
19$^{h}$23$^{m}$43$^{s}$.95, Declination
(J2000)=14${\degr}$30$\arcmin$34$\farcs$00. \textit{e8} is
$\sim$7$\arcsec$ south of the phase center. The primary beam
(field of view) of the SMA at 345 GHz is $\sim$30$\arcsec$.

Linear polarization (LP) observations using interferometer arrays
are best obtained using receivers which detect both orthogonal
circular polarizations (CP) simultaneously. However, the SMA
receivers are intrinsically linearly polarized and only one
polarization is available currently. Thus, quarter-wave plates
were installed in order to convert the LP to CP. Detailed
information of the design of the quarter-wave plates and how the
quarter-wave plates were controlled is described in Marrone et al.
(2006) and Marrone \& Rao (2008). We assume that the smearing due
to the change of the P.A.s on the time scale of 5 minutes in one
cycle of polarization measurement is negligible.

The conversion of the LP to CP is not perfect. This instrumental
polarization (also called the leakage terms)
\citep[see][]{sault96} and the bandpass were calibrated by
observing 3c454.3 for 2 hours while it was transiting in order to
get the best coverage of parallactic angles. The instrumental
polarization is $\sim$1\% for the upper sideband and $\sim$3\% for
the lower sideband before calibration, and $\sim$0.6\% after
calibration in both sidebands. The complex gains were calibrated
every 12 minutes by observing 1751+096 until it set, followed by
1925+211 for the last 3.5 hours. The absolute flux scale was
calibrated using Titan.

The data were calibrated and analyzed using the MIRIAD package.
After the standard gain calibration, self-calibration was also
performed by selecting the visibilities with uv distances longer
than 40 k$\lambda$. In order to Fourier transform the measured
visibilities to the image, the task INVERT in MIRIAD was used with
natural weighting. The Stokes $Q$ and $U$ maps are crucial for the
derivation of the polarization. We use the dirty maps of $Q$ and
$U$ to derive the polarization in order to avoid a possible bias
introduced from the CLEAN process. We applied CLEAN to the Stokes
$I$ (total intensity) map in order to reduce the sidelobes. The
presented SMA images have all been corrected for the primary beam
attenuation. The synthesized beam of the presented maps is
0$\farcs$7$\times$0$\farcs$6 with a P.A. of $-$58$\degr$. The
presented polarization vectors are gridded to a 0$\farcs$3 spacing
- which is about half of the synthesized beam FWHM - in order to
show the curvature of the B field morphology. Therefore, adjacent
polarization vectors are not formally independent within one
synthesized beam. However, as usual, relative information can be
extracted at under the synthesized beam resolution.

The Stokes $I$, $Q$ and $U$ images of the continuum are
constructed with natural weighting in order to get a better S/N
ratio for the polarization. The noise levels of the $I$, $Q$ and
$U$ images are $\sim$ 60, 4 and 4 mJy Beam$^{-1}$, respectively.
The strength ($I_{\rm p}$) and percentage ($P(\%)$) of the
linearly polarized emission are calculated from: $I_{\rm p}^{2} =
Q^{2} + U^{2} - \sigma_{Q,U}^{2}$ and $P(\%) = I_{\rm p}/I$,
respectively. The term $\sigma_{Q,U}$ is the noise level of the
Stokes $Q$ and $U$ images, and it is the bias correction due to
the positive measure of $I_{\rm p}$ (Leahy 1989; Wardle \&
Kronberg 1974). The $\sigma_{I_{\rm p}}$ is thus 4 mJy
beam$^{-1}$. To derive the polarization, the MIRIAD task IMPOL was
used. The SMA polarization vectors presented are above
3$\sigma_{I_{\rm p}}$ in red segments and between 2 to 3
$\sigma_{I_{\rm p}}$ in black segments.

\section{Results}
The 870 $\mu$m continuum emission and its polarized components
were detected (Figure 2; Table 1 \& 2). The results are presented
in this section.

\subsection{Continuum Emission}

In e2, a compact 870 $\mu$m continuum emission structure with a
radius of $\sim$ 1$\arcsec$ (0.03 pc) is centered at
$\sim$0$\farcs$7 east of \textit{e2}. Extending to the north-west
of this compact emission, a fainter structure with an overall
length of $\sim$ 2$\arcsec$ (0.07 pc) is detected. The H$_{2}$O
\citep{genzel81} and (J,K)=(9,6) NH$_{3}$ \citep{pratap91} masers
are located in this north-west extension, $\sim$ 2$\arcsec$ away
from the continuum peak. Associated with the continuum peak, there
are OH masers detected within 0$\farcs$5 to the east and
$\sim$1$\arcsec$ to the south of \textit{e2}
\citep{gaume87,fish06}, suggesting that it is an active star
forming site.

In e8, the 870 $\mu$m continuum peak is centered at 0$\farcs$3
west of \textit{e8}. \textit{e4}, \textit{e1} and \textit{e3} are
at the periphery of the 870 $\mu$m continuum emission. There is an
extension toward the south-west with an overall length of
$\sim$3$\arcsec$. Associated with \textit{e8}, an NH$_{3}$ maser
spot was detected 0$\farcs$8 south of the 870 $\mu$m continuum
peak by Pratap et al. (1991). The OH \citep{gaume87} and H$_{2}$O
\citep{genzel81} masers are also associated with \textit{e8} and
the 870 $\mu$m continuum peak, suggesting again that this is an
active star formation site.

When fitted with a Gaussian, the deconvolved size of the 870
$\mu$m emission in e2 is 0$\farcs$9$\times$0$\farcs$8, slightly
larger than the synthesized beam, and therefore, the e2 core has
been resolved. For e8, the deconvolved size is
0$\farcs$9$\times$0$\farcs$3 with a P.A. of 12$\degr$. Therefore,
e8 has been resolved along the major axis of the dust ridge but
not along the minor axis. In both e2 and e8, the 870 $\mu$m
continuum emissions are associated with the NH$_{3}$ cores
\citep{ho83,zhang97}, suggesting that they are also tracing the
dense regions.

The measured 870 $\mu$m flux densities within the upper and lower
boxes in Figure 2(b), associated with e2 and e8, are 9.3 and 4.0
Jy, respectively. The flux densities of the free-free continuum
$F_{ff}$ at 1.3 cm in e2 and e8 are 300 mJy \citep{gaume93} and 17
mJy \citep{zhang97}, respectively. In order to estimate the
$F_{ff}$ contribution at 870 $\mu$m, we extrapolate from 1.3 cm,
assuming $F_{ff}$ $\propto$ $\nu^{-0.1}$. Although this assumption
of optically thin emission is crude, it has been shown that the
resultant $F_{ff}$ roughly agrees (within a factor of 3) with the
estimate from the radio recombination line at 2 mm
\citep{zhang98}, suggesting that the assumed $F_{ff}$ $\propto$
$\nu^{-0.1}$ is reasonable. The extrapolated $F_{ff}$ at 870
$\mu$m is $\sim$230 and 13 mJy for e2 and e8, respectively. As
compared to the 870 $\mu$m flux densities, $F_{ff}$ contributes
$\sim$2\% for the e2 region and 0.3\% for the e8 region.
Therefore, the 870 $\mu$m continuum is dominated by dust emission.
Hereafter, the structures traced by the 870 $\mu$m emission in the
e2 and e8 regions are named as e2 dust ridge and e8 dust ridge,
respectively.

Assuming a dust temperature of 100 K \citep{zhang98}, a dust grain
emissivity \textit{Q($\lambda$)} $\propto$ $\lambda^{-\beta}$ with
$\beta = 1$, and the normal gas to dust ratio of 100, we estimate
gas masses $M_{\rm gas}$ of 245 and 106 M$_{\sun}$ for the e2 and
e8 dust ridges, respectively (cf. Tang et al. 2009). Note that the
$M_{\rm gas}$ given here is highly affected by the assumed
$\beta$. If the assumed $\beta$ is 2, the estimated $M_{\rm gas}$
will be 14 times larger. Assuming the extents along the line of
sight are equal to the diameters of the emission area in the e2
and e8 dust ridges, the average gas number densities $n_{\rm
H_{2}}$ are 3.4$\times$10$^{6}$ and 2.2$\times$10$^{6}$ cm$^{-3}$,
respectively. By using the same equation and the same assumed
values of $\beta$ and dust temperature, the $M_{\rm gas}$
estimated from the 2 mm dust continuum \citep{zhang98} for the e2
and e8 dust ridges are 1100 and 590 M$_{\sun}$, respectively. The
difference in the estimated $M_{\rm gas}$ at 2 mm and 870 $\mu$m
is most likely due to the missing flux from the extended
component, which is not recovered with our SMA observations. In
comparison, with the same assumptions, the $M_{\rm gas}$ of the
envelope is 1834 M$_{\sun}$ as traced at 1.3 mm by BIMA
\citep{lai01}. The $M_{\rm gas}$ associated with the e2 and e8
dust ridges recovered with our SMA observations is $\sim$ 19\% of
the $M_{\rm gas}$ in the envelope.

The main conclusion from the dust continuum data is that the associated
mass is large. The morphology of the dust continuum is elongated.
The positional offsets between the various embedded sources are
significant, such as between the positions of the 870 $\mu$m peaks and the
UCHII regions. These results are consistent with the formation of
a cluster of stars.


\subsection{Dust Polarization}

The polarization in the e2 and e8 dust ridges is detected and
resolved (Figure 2 (c) and (d)). Throughout the paper, P.A. is
defined from the north to the east. In the e2 dust ridge, the bulk
of the polarization vectors form a ring around the 870 $\mu$m peak
with a radius of $\sim$1$\arcsec$ and with the geometric center
near the continuum peak instead of \textit{e2}. In the north-west
extension of the dust ridge, the polarization appears to be
perpendicular to the major axis of the extension.

The e8 dust ridge is $\sim$7$\arcsec$ away from the phase center.
Even though the antenna response is 15\% less efficient than at
the phase center, the polarization revealed is clearly also not as
uniform as previously seen with BIMA. The polarization vectors
again form a ring like structure around the continuum peak.  The
polarization is weaker in e8 with more vectors between 2 to 3
$\sigma_{I_{\rm p}}$.

In comparison, the polarization in the envelope of the e2 and e8
regions, as revealed with an angular resolution of 3$\arcsec$ (0.1
pc) with BIMA, shows a relatively uniform distribution in P.A. and
therefore, a fairly uniform B$_{\bot}$ field at 1.3 mm
\citep[Figure 2(a);][]{lai01}. In their results, the polarization
in the e2 region is weak and resolved into e2 main and e2 pol NW,
named in the same paper, according to the P.A. of the polarization
vectors. The component e2 pol NW is at 3$\arcsec$ to the
north-west of \textit{e2}. There is a gap where no polarized
emission is detected between \textit{e2} and e2 pol NW. In the e8
region, the polarization in the BIMA results is nearly uniform
with a decrease in polarization percentage near the peak position.

In order to test if the differences in polarization properties
from SMA and BIMA are due to their different angular resolutions,
we smoothed our SMA results to the BIMA resolution, as shown in
Figure 3. Wherever the polarized emission was both detected at 1.3
mm and 870 $\mu$m, the resultant P.A.s of the polarization
differed by $\sim$ 30$\degr$ on average. This significant
difference can be due to the different sampling of the
visibilities, which are in the range of 6 to 170 k$\lambda$
($\lambda$=1.3 mm) for the BIMA and in the range of 30 to 262
k$\lambda$ ($\lambda$=870 $\mu$m) for the SMA. Specifically, the
SMA filtered out the more extended and uniform component which is
larger than 8$\arcsec$. At the same angular resolution, the
derived global B$_{\bot}$ field directions in e2 and in e8 are
therefore consistent in the regions where both the SMA and BIMA
have polarization detections. Most importantly, the smoothed SMA
polarization map shows that the polarization percentage has
decreased significantly, especially near the continuum peak
positions, where the field geometry is more complex at the
resolution of 0$\farcs$7. This demonstrates that the low
polarization percentage at the emission peaks is due to the
limited angular resolution when a more complex underlying B field
morphology has not been resolved. This effect can also be due to
the decrease of the alignment efficiency of the dust grains in
denser regions (Lazarian \& Hoang 2007) or due to geometrical
effects, such as the differences in the viewing angles
(Gon\c{c}alves et al. 2005). However, in this case, the complex B
field structure is the dominant effect.

The polarization percentage $P(\%)$ decreases with increasing
continuum intensity $I$ in both e2 and e8 even for the higher
resolution SMA results (Figure 4). Since the BIMA results come
from a resolution effect, the same might be true for the SMA
results at the emission peaks.  Away from the emission peaks, the
general increase in $P(\%)$ is somewhat misleading. Figure 2(a)
shows that this effect is not symmetrical on either side of the
elongated envelope, i.e. the $P(\%)$ differs with positions on the
same contour level of $I$. This is reflected by the large
dispersion in $P(\%)$ at any value of $I/I_{max}$. Several
effects, including B field geometry related to the line of sight,
need to be disentangled. That the $P(\%$) ranges mainly between
1\% to 10\% (Figure 4), seems to agree with the model of grain
growth in the dense regions where grain alignments are via
radiative torques \citep[see Figure 11 in][]{pelkonen09}. However,
based on our results, the effects of angular resolution and
geometry must first be taken into account.


\section{Discussion}








\subsection{Hourglass B field Morphology inside the e2 dust ridge?}
The inferred B field in the e2 dust ridge exhibits a complex but
organized morphology (Figure 5). We have tested the hypothesis of
the measured B field being radial, and have shown quantitatively
the preference of a non-radial field at a high significance level
(see Appendix). There are positions where no polarized emission
(depolarization) is detected, extending along a P.A. $\sim$
60$\degr$ across the 870 $\mu$m peak (color scale in Figure 2(c)).
The existence of non-radial field lines together with the
depolarized zones are in favor of an hourglass field morphology.
Along the extension of the dust ridge toward the north-west, the B
field lines are approximately parallel to the major axis, which is
consistent with the BIMA measurement at 1.3 mm. These lines are
radial-like, but the complex structure in the north-west could
belong to another embedded source as possibly indicated by the
masers.

Associated with \textit{e2}, organized motions in the ionized gas
have been revealed with the H53$\alpha$ radio recombination line
\citep{keto08}, with the maximum velocity gradient along P.A.
$\sim$ 60$\degr$. These authors interpret this gradient as a
supporting evidence for an accretion flow along a dense flattened
structure, where the detected motion tracks the ionized particles
on the surface of the dense midplane. Both the infall and rotation
near \textit{e2} have also been detected in several molecular
lines \citep{ho96,zhang97,zhang98}.
As discussed in Keto \& Klaassen (2008), this H53$\alpha$
accretion flow in the direction of P.A. $\sim$ 60$\degr$ might
drive the molecular outflow at P.A. $\sim -$ 20$\degr$ as traced
by the CO 2-1 line. The argument is based on the hypothesis that
if the massive star formation process is similar to the low mass
case, the bipolar outflow should be along the rotation axis. The
linearly distributed H$_{2}$O and OH masers in the W51 e2 region
could trace an outflow \citep[Figure 19(c),][]{debuizer05}, as
identified with the CO 2-1 line. Although the determined direction
may be highly uncertain, the rotation in NH$_{3}$ (3,3) is more
clearly revealed along PA=135$\degr$ \citep[Figure 7
in][]{zhang97} and in CH$_{3}$CN along PA=110$\degr$
\citep{zhang98}, which seems inconsistent with the gradient
detected with the H53$\alpha$ ionized flow. This might indicate
that the revealed kinematics based on different lines may be from
multiple embedded sources. Higher spatial resolution kinematic
studies with hot core molecular lines will be helpful for
deciphering the underlying structures.




The B field appears to be hourglass-like near \textit{e2}, with
the field lines pinched along the plane of the proposed
H53$\alpha$ accretion disk. If the B field lines are frozen into
the ionized material, the field lines will be tangled along with
the rotation and infall motions.  The revealed depolarization
might then result from the more complex underlying B field. We
note that the field lines seem to go to the core with an
essentially radial pattern, and therefore, leading to a sharp
pinched angle in the hourglass. In contrast, the low mass case
\citep{girart06} shows a wider and smoother pinched angle. We
speculate that a larger infall momentum and a larger differential
rotation \citep{zhang98} might drag the field lines along and
result in a narrower pinched angle in the projected plane.
Projection of a nearly pole-on hourglass-like morphology possibly
also leads to similar signatures. In any case, the scenario of
material accreting through a disk as proposed by Keto and Klaassen
(2008) is supported by our inferred B field morphology.

\subsection{Hourglass B field Morphology in the e8 dust ridge?}

Along the e8 dust ridge, the B field also shows a systematic
deviation from the larger scale (0.5 pc) B field revealed by
BIMA. This can again be explained by the field lines being dragged
along with the accretion toward \textit{e8}. In this case, the
revealed B field appears to be part of an hourglass structure on a
larger scale of 4$\arcsec$ ($\sim$0.08 pc) (Figure 5(b)), with its
pinched direction parallel to the dust ridge. Centered on the e8
continuum peak, a compact hourglass structure would be more
convincing except for the field lines to the north.  There are
H$_2$O masers north of the e8 continuum peak, and another embedded
source may be indicated.  This could explain the incomplete
hourglass structure here.

As in the case of e2, a zone of depolarization seems to be present
at the continuum peak, along the north-south direction.  This is
consistent with the pinch direction of the hourglass-like
morphology being along the elongated e8 dust ridge. Rotation
associated with the e8 collapsing core was detected in the
direction of P.A. $\sim$ 156$\degr$ ($-$24$\degr$) with CH$_{3}$CN
\citep{zhang98}. In this scenario, the pinch direction of the
hourglass-like B field is parallel to the plane of rotation. The
rotation axis of the e8 collapsing core is then almost parallel to
the B field threading the 870 $\mu$m dust ridge. Note that the
rotation direction as traced in CH$_{3}$CN is still uncertain
\citep{zhang98}. An accurate determination of the plane of
rotation associated with the e8 collapsing core is needed to test
if the larger scale B field controls the direction of accretion.



Although the plane of accretion (or the pinched angle of the
hourglass) cannot be determined with certainty, the collapse
signature was detected toward \textit{e8}
\citep{ho96,zhang97,zhang98}, consistent with the possible
hourglass-like B field morphology. Furthermore, this collapsing
core is inside the 0.08 pc scale dust ridge, as revealed with the
0$\farcs$7 angular resolution B field morphology. Based on this
morphology and the presence of \textit{e4}, \textit{e8},
\textit{e1}, and \textit{e3}, we suggest that the star formation
process involves different stages of fragmentation, proceeding at
different evolutionary timescales.


\subsection{Estimate of the Strength of the B field}
The B field strength can be estimated by comparing the
gravitational force $f_{\rm G}$ with the B field tension $f_{\rm
B}$ following Dotson (1996) and Schleuning (1998). The value of
$f_{\rm G}$ at a distance $R_{\rm G}$ away from the center is
given by

\begin{equation}\label{1}
    f_{\rm G} = \frac{G\hspace{0.2em}M_{\rm R} \hspace{0.2em}\rho}{R_{\rm G}^2} =
    5\times10^{-26}\frac{M_{\rm R}}{100\rm M_{\sun}} \frac{n_{\rm H_{\rm
    2}}}{10^{5}\hspace{0.2em}\rm cm^{-3}}(\frac{R_{\rm G}}{0.1\hspace{0.2em}\rm pc})^{-2}
    \hspace{1em}\frac{{\rm dyne}}{{\rm cm^3}},
\end{equation}
where $M_{\rm R}$ refers to the gas mass enclosed within a radius
$R_{\rm G}$, $\rho$ is the mass density at $R_{\rm G}$, and
$n_{\rm H_{\rm 2}}$ is the gas volume number density. The $f_{\rm
B}$ can be given by
\begin{equation}\label{2}
    f_{\rm B} = \frac{1}{4\pi} \vec{B}\cdot\vec{\bigtriangledown} \vec{B} \sim \frac{B^2}{4\pi
    R_{\rm B}} = 5\times 10^{-26}(\frac{B}{\rm mG})^2(\frac{R_{\rm B}}{0.5 \hspace{0.2em}\rm pc})^{-1}
    \hspace{1em}\frac{{\rm dyne}}{{\rm cm^3}},
\end{equation}
where $R_{\rm B}$ is the radius of a magnetic flux tube, and $B$
is the B field strength. Since the e2 and e8 cores are known to be
in a collapse stage, we conclude that $f_{\rm G}$ $>$ $f_{\rm B}$.
An upper limit of $B$ can then be derived. For the e2 collapsing
core, $M_{\rm R}$ is estimated to be 220M$_{\sun}$ based on the
870 $\mu$m flux density within a radius of 1$\arcsec$ of the
continuum peak. This is consistent with the proposed self
gravitating mass of $>$160 M$_{\sun}$ \citep{ho96} based on the
kinematics of the NH$_{3}$ lines. $R_{\rm G}$ is 1$\arcsec$ (0.034
pc), and the mean $n_{\rm H_{\rm 2}}$ within $R_{\rm G}$ is
2.7$\times$10$^{7}$ cm$^{-3}$. Assuming $R_{\rm G}$ = $R_{\rm B}$
$\simeq$ 0.034 pc, the B field strength in the e2 core is
therefore $<$ 19 mG. In the e8 collapsing core, $M_{\rm R}$ is 94
M$_{\sun}$, and $n_{\rm H_{\rm 2}}$ is 1.2$\times$10$^{7}$
cm$^{-3}$ within a radius of 1$\arcsec$ centered on the peak
position. The B field strength in the e8 core is therefore $<$ 8
mG. Both of the upper limits of $B$ are consistent with the lower
limit of the larger scale B$_{\bot}$ field strength of 1 mG
\citep{lai01} estimated from the method proposed by Chandrasekhar
\& Fermi (1953).
\subsection{Characteristic Length Scales} To analyze the interactions
between B field, gravitational force and thermal force in star
forming sites, we further calculate the following three length
scales following Mouschovias (1991): First, the interplay between
ambipolar diffusion and Alfv$\acute{e}$n waves is characterized by
the Alfv$\acute{e}$n length scale $\lambda_{\rm A}$. Second, the
interplay between gravitational and thermal pressure forces is
characterized by the thermal Jeans length scale $\lambda_{\rm
T,cr}$, following Bonnor (1956) and Ebert (1955; 1957). Third, the
interplay between magnetic and gravitational forces is
characterized by the critical magnetic length scale $\lambda_{\rm
M,cr}$. They can be calculated using the following equations:
\begin{equation}\label{A}
    \lambda_{\rm A} = 8 \frac{B}{\rm mG}(\frac{n_{\rm H_{\rm 2}}}{{10^{6}}\hspace{0.2em}\rm
    cm^{-3}})^{-1}
    (\frac{K}{3\times10^{-3}})^{-1}
    \hspace{1em} {\rm mpc},
\end{equation}
\begin{equation}\label{T}
    \lambda_{\rm T,cr} = 31 \sqrt{\frac{T}{100\hspace{0.2em}\rm K}
    (\frac{{n_{\rm H_{\rm 2}}}}{10^{6}\hspace{0.2em}\rm cm^{-3}})^{-1}}
    \hspace{1em}{\rm mpc},
\end{equation}
and
\begin{equation}\label{M_cr}
\lambda_{\rm M,cr} = 36\frac{B}{\rm mG}(\frac{n_{\rm H_{\rm
2}}}{10^6\hspace{0.2em}\rm cm^{-3}})^{-1} \hspace{1em} {\rm mpc}.
\end{equation}
Here, the parameter $k$ (Eq. (6f) in Mouschovias 1991), related to
the mean collision time between an ionized and a neutral particle,
is assumed to be 0.5 when we derive Eq. (3), which is within the
most likely range given in the reference in their paper. The
factor $K$ is related to the cosmic ionization rate. We assume
\textit{K} = 3$\times$10$^{-3}$ \citep{mouschovias91} following
the ionization rate calculated by Nakano (1979). With the assumed
\textit{k} and \textit{K}, the estimated fractional ionization
rate is 3$\times$10$^{-9}$ for a number density of 10$^{7}$
cm$^{-3}$, which seems to be reasonable. $T$ is the gas
temperature, and $n_{\rm H_{\rm 2}}$ is the gas volume number
density. $B$ is the B field strength. Note that \textit{T} is
assumed to be 100 K in both the e2 and e8 dust ridges based on the
analysis of the hot core lines by Zhang, Ho \& Ohashi (1997).
Since these natural length scales depend on $n_{\rm H_{\rm 2}}$
and $B$, they are calculated separately based on the detected
continuum emission with the same assumption as in \S 3.1. In the
1.3 mm envelope, the $n_{\rm H_{\rm 2}}$ is the mean number
density within a best-fit Gaussian centered on the peak, and $B$
is the lower limit of 1 mG. In the 870 $\mu$m dust ridges, $n_{\rm
H_{\rm 2}}$ is calculated within a radius of 1$\arcsec$, and $B$
is the upper limit calculated in \S 4.3. The calculated natural
length scales in the e2 and e8 cores are listed in Table 3.


The physical meaning of these length scales are explained clearly
in Mouschovias (1991) and references therein. $\lambda_{\rm A}$
gives the lower limit of the scale at which the B field can
sustain the structure. At the scale R$<\lambda_{\rm A}$, the
ambipolar diffusion between neutral and ionized particles is more
efficient and the Alfv$\acute{e}$nic motion is less important.
$\lambda_{\rm T,cr}$ gives the scale where the gravitational force
is equal to the thermal pressure. If an object has a size scale R
$>\lambda_{\rm T,cr}$, gravity can overwhelm the thermal pressure
and collapse will start. $\lambda_{\rm M,cr}$ gives the upper
limit of the scale where the cloud can be magnetically supported
along the B field direction. In a region R$>\lambda_{\rm M,cr}$,
there will be enough mass and therefore the material can collapse
if there are no other supporting forces.


\subsubsection{Correlation with the SMA e2 dust ridge}

In the e2 dust ridge, the revealed B$_{\bot}$ morphology is
clearly pinched with a radius of $\sim$ 0$\farcs$8 near
\textit{e2}, comparable to the radius of the proposed rotating
flattened structure of 1$\arcsec$ \citep{zhang97,zhang98} and the
proposed ionized accretion disk of $\sim$ 0$\farcs$5
\citep{keto08}. The derived $\lambda_{\rm T,cr}$ is $\sim$
0$\farcs$2, suggesting that at $\sim$ 1$\arcsec$, gravity will
easily overcome the thermal pressure support if there are no other
supporting forces. The scale where ambipolar diffusion starts to
take place ($\lambda_{\rm A}$) is $\sim$ 0$\farcs$2 for the e2
dust ridge, consistent with the observed pinched B$_{\bot}$ field
lines. Note that the revealed width of the depolarization zone
near \textit{e2} is narrow ($<$ 0$\farcs$5), which is smaller than
our synthesized beam. Higher angular resolution measurements with
at least 0$\farcs$3 resolution are needed to discriminate whether
the depolarization is due to ambipolar diffusion, inefficient
grain alignment or other mechanisms, such as geometrical effects.
The calculated scale where the B field can sustain the structure
against gravitational collapsing ($\lambda_{\rm M,cr}$) is
0$\farcs$8 along the field line. However, it is difficult to
compare $\lambda_{\rm M,cr}$ with the scale associated with the
dust ridge, because the large scale (0.5 pc) B$_{\bot}$ field is
twisted by $\sim$45$\degr$ (Figure 2(a)) at 3$\arcsec$ resolution.
Observations with visibilities at both shorter and longer uv
ranges are needed in order to link the B$_{\bot}$ in the core with
the field in the envelope at the same wavelength. Note that the
weak constraints on $\lambda_{\rm A}$ and $\lambda_{\rm M, cr}$
result from the large range of possible B field values. 


\subsubsection{Correlation with the SMA e8 dust ridge}
The dust continuum emission appears to be ridge-like, and the
minor axis of the e8 dust ridge is approximately parallel to the
0.5 pc B$_{\bot}$ field direction. 
The deconvolved length of the e8 dust ridge along the minor axis
is barely resolved, and we adopt an upper limit to the radius of
0$\farcs$3 along the minor axis. This is consistent with the
estimated $\lambda_{\rm M,cr}$ $<$ 0$\farcs$7, and the estimated
$\lambda_{\rm T,cr}$ of 0$\farcs$3. This suggests that thermal
pressure is significant as compared to gravity and field tension
at this scale. Along the major axis, the deconvolved size is
0$\farcs$9, which is larger than $\lambda_{\rm T,cr}$.
Furthermore, ambipolar diffusion is expected to dominate at the
scale $\lambda_{\rm A}$ $<$ 0$\farcs$2 in e8. Hence we expect
collapse and fragmentation to occur along the ridge. This is
consistent with the revealed hourglass-like B field morphology
associated with the e8 collapsing core at 0$\farcs$7 resolution
and the smooth B field morphology in the envelope at 3$\arcsec$
resolution. These results in the e8 core seem to be consistent
with the ambipolar diffusion model \citep{mouschovias91a} and
suggest that the formation of the dust ridge is influenced by the
B field in the envelope.

\subsubsection{Connection between the BIMA 0.5 pc scale envelope and the SMA dust ridges}

The main difference between the 0.5 pc envelope and the dust
ridges is in the large contrast in the derived value for $n_{\rm
H_{\rm 2}}$. Nevertheless, the agreement in the estimates of
$\lambda_{\rm A}$ and $\lambda_{\rm M,cr}$ for the envelope and
the ridge is very good, because of the lower estimates of $B$ in
the envelope. This implies that the B field support is adequate
till the scale of 0$\farcs$2$-$1$\arcsec$. The derived value of
$\lambda_{\rm T,cr}$ is larger for the envelope than for the
ridge, but still smaller than the measured size of the envelope.
This is also consistent with the proposed scenario that the B
field provides the support against gravity at the 0.5 pc scale.

\subsection{Role of B$_{\bot}$ from Envelope (0.5 pc) to Collapsing Cores (0.02 pc)}
The structure of the B field varies with different size scales
(Figure 6).

At the 0$\farcs$7 (0.02 pc) scale, the B field in the W51 e2 and
e8 cores is not uniform. In both the e2 and e8 cores, the
hourglass-like morphology suggests that at the 0.02 pc scale,
gravity dominates over the B field. Our estimated collapsing
$M_{\rm gas}$ is on the order of 100 M$_{\sun}$ (see \S 4.3) in
both the e2 and e8 cores, which is roughly consistent with $M_{\rm
gas}$ estimated in the local collapsing scenario
\citep{ho96,sollins04}. The cores are therefore in the
supercritical phase.

At the 0.5 pc scale, the B field in the envelope is uniform
throughout the e2 and e8 dust ridges, except to the north-west of
e2. Due to the small dispersion of the measured P.A.s of the
polarization in the envelope, the B field is suggested to dominate
over the turbulence with a strength of $\gtrsim$1 mG by Lai et al.
(2001). We further suggest that the envelope is subcritical. The
reasons are: 1. B field dominates over turbulence in the envelope.
2. B field is apparently sufficient to support the envelope
against gravity (\S 4.4.3). 3. The collapse is apparent only
locally in e2 and e8 with $M_{\rm gas}$ in the order of 100
M$_{\sun}$, so that the envelope is stable.


Based on the MHD simulations \citep{klessen00}, turbulent motions
(cf. Mac Low \& Klessen 2004) can produce an elongated envelope
and can sustain the envelope from collapse. However, such
elongated structures would not exhibit a preferred alignment with
the B field \citep{heitsch01}, as it is seen here.
The measured B field structures from 0.02 pc to 0.5 pc do not show
an obvious necessity for turbulent support.  Instead, the cloud
morphology, sizescale, and B field geometry are consistent with
\textit{magnetic fragmentation} via ambipolar diffusion
\citep{mouschovias91a,lizano89,shu04}. 




\subsection{Comparison with other star formation sites}
Because the collapse signatures in the W51 e2 and e8 regions are
clearly detected, and the UCHII regions are still relatively
compact and weak, they are in an earlier evolutionary stage as
compared with the other massive star formation sites, such as
G5.89$-$0.39 (Tang et al. 2009). In G5.89-0.39, the B field is
suggested to be overwhelmed by the turbulent motions from the
UCHII expansion and the molecular outflows. A more complex
B$_{\bot}$ morphology is detected (Tang et al. 2009) with a
spatial resolution of 0.02 pc. In contrast to G5.89$-0.39$, the
B$_{\bot}$ morphologies are hourglass-like in both of the W51 e2
and e8 regions at the 0.02 pc spatial resolution, but much
smoother at the 0.5 pc scale. This comparison of G5.89-0.39 and
W51 e2/e8 with the same spatial resolution indicates that the role
of the B field varies with the evolutionary stages of the central
sources.

In the low mass star formation region NGC 1333 IRAS 4A, the
hourglass shape is observed on a scale of 2400 AU (0.01 pc) and
$M_{\rm gas}$ is $\sim$1.2M$_{\sun}$ (Girart, Rao, \& Marrone
2006). In comparison, the hourglass B$_{\bot}$ structure detected
in W51 e2 is on the scale of $\sim$ 0.03 pc (1$\arcsec$), and the
mass involved is $\sim$ 200 M$_{\sun}$. The consistency of the
directions of the ionized flow and the pinched field further
suggests that the stars are formed with similar mechanisms, i.e.
material is accreted through a flattened structure. The difference
is that the scale and the mass involved are much larger in the
massive star forming regions. At the time of publication of this
paper, additional observational evidence of an hourglass B field
morphology in the massive star forming core G31.41+0.31 is
presented in Girart et al. (2009). This further supports the
proposed similar formation mechanism as in the low mass case.

However, the massive star forming sites are much further away than
the low mass regions in general. For example, W51 e2 is at 7 kpc,
$\sim$ 23 times further away than NGC 1333 IRAS 4A. Observations
with higher spatial resolution are thus needed for the massive
star forming regions. The closest massive star forming site is
Orion BN/KL. Source I in Orion BN/KL is suggested to be in an
early stage of massive star formation due to the weak and compact
free-free emission \citep{plambeck95}. Source I is also suggested
to harbor an ionized accretion disk \citep{reid07}. The B$_{\bot}$
at 0.5 pc starts to exhibit a larger scale hourglass-like
morphology \citep{schleuning98}. The existence of the ionized
disk, the uniform large scale B$_{\bot}$ geometry and the compact
free-free source all suggest that it is an analog of W51 e2/e8,
but at a much closer distance. The comparison between w51 e2/e8
and Source I in Orion BN/KL may provide a clue of the B$_{\bot}$
morphology at even higher physical resolutions.

In W51 e2/e8, we found two supercritical cores at 0.03 pc within a
subcritical envelope at 0.5 pc (\S 4.5). At a larger scale, DR21
MAIN is suggested to have started undergoing a gravitational
collapse in the central part of the cloud \citep{kirby09}.
Suggested by the same author, in the outer part ($\sim$ 1 pc away
from the center), the cloud is still magnetically supported. That
the collapsing cores formed inside a magnetically supported cloud
as in W51 e2/e8, is therefore not a special case. As indicated by
Vaillancourt (2009), observations of B field at the larger scales
are needed to test the magnetically controlled star formation
process. Shorter spacing visibilities of the SMA are needed to
directly compare the field morphology of the cores with the
envelope at the same wavelength of 870 $\mu$m. In this paper, we
have demonstrated that smaller scale B$_{\bot}$ maps provide
crucial information about the B$_{\bot}$ field by resolving the
star forming cores and by linking the field morphology with the
kinematics of the molecular cloud.



\section{Conclusion and Summary}
To study the role of the B field in the star forming cores, we
have observed and analyzed the B$_{\bot}$ morphologies, inferred
from the linearly polarized dust continuum emission, in the
massive star forming site W51 e2/e8 by using the Submillimeter
Array (SMA). We further compare the B$_{\bot}$ morphologies in the
dust ridges with the one in the envelope. Three different natural
length scales, namely the Alfv\'{e}n length scale $\lambda_{\rm
A}$, thermal Jeans lengths scale $\lambda_{\rm T,cr}$, and the
magnetic length scale $\lambda_{\rm M,cr}$, are calculated and
compared to the dust ridges and envelope. Here are the summary and
conclusion.

\begin{enumerate}
    \item The 870 $\mu$m continuum and its polarization in W51 e2
    and e8 are resolved with an angular resolution of 0$\farcs$7 as observed with the SMA.
    The polarization in both e2 and e8 exhibit complex structures. In comparison,
    the polarization at 1.3 mm observed at 3$\arcsec$ resolution with the BIMA
    revealed a uniform morphology across e2 and e8, with almost no polarization
 detected near the peak position in e2 \citep{lai01}.
    We conclude that low or no polarization near the emission peaks heretofore
    seen in the star formation regions
    is likely due to a more complex underlying B$_{\bot}$ morphology (\S 3.2).

    \item In the e2 dust ridge, the inferred B$_{\bot}$ morphology is hourglass-like
    near the collapsing core, with its pinched direction parallel to the direction of the ionized
    accretion flow as traced by H53$\alpha$ \citep{keto08}.
The B$_{\bot}$ here shows a similar morphology as in the low mass
star formation case NGC 1333 IRAS 4A. However, the mass included
in this core is $\sim$200 times larger. This result shows that the
B field in the e2 collapsing core plays a similar role as in the
low mass star formation regions (\S 4.1). Higher angular
resolution observations are required to test if the hourglass-like
B$_{\bot}$ morphology is preserved at smaller scale. 

\item The e8 dust ridge is perpendicular to the 0.5 pc scale
B$_{\bot}$ direction as revealed with the BIMA, which suggests
that the B field at 0.5 pc controls the forming process of the
dust ridge. The B$_{\bot}$ along the dust ridge exhibits a
systematic deviation from the B$_{\bot}$ at 0.5 pc scale.
Associated with the e8 collapsing core, the hourglass B$_{\bot}$
morphology is more clearly detected, suggesting that the
collapsing core is formed locally inside a flattened structure (\S
4.2).

    \item The exhibited hourglass-like B$_{\bot}$ morphologies in the e2 and possibly the e8
    dust ridges are consistent with the proposed local collapse \citep{ho96,sollins04}
    instead of the global collapse \citep{rudolph90}. This indicates that both the e2 and e8 cores
    are in a supercritical stage (gravity dominating over the B field) at the 0.02 pc scale.
    In contrast,
    the B field morphology of the 0.5 pc envelope is uniform across the e2 and e8 dust ridges
    and strong ($\geq$1 mG). We further propose that the B field
    in the 0.5 pc scale envelope is subcritical (B field dominating the
    gravity). That the supercritical cores formed inside a subcritical
    envelope seems to support a \textit{magnetic fragmentation}
    scenario \citep{mouschovias91a,lizano89,shu04}, suggesting that ambipolar diffusion
    plays a key role in the evolution of the envelope at this
    stage (\S 4.5).

\end{enumerate}

\textit{Acknowledgement} The authors are grateful for the
anonymous referee's comments, which helped to improve the
manuscript.

\section*{Appendix}

In order to test how significantly the measured B field deviates
from a purely radial field, we have analyzed the differences,
$\delta$, between position angles (P.A.s) of the measured B field
lines and the corresponding hypothetical radial field lines. The
radial field lines are derived from their relative positions
(center of each P.A.) with respect to the origin, which is defined
as the 870 $\mu$m peak in e2. Due to the limited detected data
points, we do not apply this statistical analysis to e8. All the
data considered are above 3 $\sigma_{I_{\rm p}}$. We have excluded
the 6 data points in the north-west extension, which correspond to
e2 pol NW. The distribution of $\delta$ apparently deviates from a
Gaussian (Figure A1(Left-panel)). When, nevertheless enforcing a
Gaussian fitting, the derived mean $\mu$ is 1.6$\degr$ and the
standard deviation $\sigma$ is 14$\degr$, which is larger than the
mean measurement uncertainty $\sigma_{\rm mean, PA}$ of
6.9$\degr$.

To quantify the deviation from a Gaussian, we further apply a
Kolmogorov-Smirnov (KS) test to $\delta$. As null hypothesis we
assume that the distribution of $\delta$ is normal with
$\sigma_{\rm mean, PA}$ and $\mu$ = 0$\degr$. This mimics an
observation of a radial field with our observed measurement
uncertainty $\sigma_{\rm mean, PA}$. The measurement uncertainties
in each P.A. are propagated with a Monte-Carlo (MC) simulation
when deriving $\delta$ and applying the KS test. In the MC
simulation we also allow for a shift of $\pm$0$\farcs$1 of the
origin of the radial field. Figure A1 (Right-panel) shows the
cumulative distributions which are used for the statistic measure
in the KS test. As a result, the probability of the measured field
being radial is 20\%. For the $\pm 1 \sigma_{\rm mean, PA}$ error
bounds, the probability of being radial is less than 5\%.
Therefore, the null hypothesis can be rejected. Our test favors
the existence of a non-radial B field.

$\delta$ as a function of P.A. is presented in Figure A2. If we
separate the segments according to the depolarization zones
(marked as arrows in Figure A2), there are always segments with
$\delta$ larger than 3 $\sigma_{\rm mean, PA}$ in each zone. This
suggests that deviations are not prevalent in certain directions,
but rather grouped together and interleaved with depolarization
zones, as we would expect from an hourglass field morphology.

\begin{deluxetable}{cccccc}

\tablecaption{SMA dust polarization at 870$\mu$m in e2}
\tablewidth{0pt} \tablehead{ \colhead{$\triangle$x} &
\colhead{$\triangle$y} & \colhead{$I$}&\colhead{$P(\%)$} &
\colhead{$I_{\rm p}$} & \colhead{$\phi(\degr)$}}

\startdata

-1.2    &   2.4 &   0.32    &   5.9 $\pm$   1.1 &   19  &   66  $\pm$   6   \\
-1.5    &   2.4 &   0.29    &   6.6 $\pm$   1.3 &   19  &   61  $\pm$   6   \\
-0.6    &   2.1 &   0.24    &   7.4 $\pm$   1.5 &   18  &   -71 $\pm$   6   \\
-0.9    &   2.1 &   0.36    &   4.4 $\pm$   1.0   &   16  &   -75 $\pm$   7   \\
0.3 &   1.8 &   0.31    &   4.8 $\pm$   1.2 &   15  &   -75 $\pm$   7   \\
0.0   &   1.8 &   0.45    &   4.2 $\pm$   0.8 &   19  &   -82 $\pm$   6   \\
-0.3    &   1.8 &   0.41    &   4.9 $\pm$   0.9 &   20  &   -89 $\pm$   6   \\
-0.6    &   1.8 &   0.43    &   3.7 $\pm$   0.9 &   16  &   -84 $\pm$   7   \\
-0.9    &   1.8 &   0.43    &   3.0   $\pm$   0.9 &   13  &   -70 $\pm$   8   \\
0.6 &   1.5 &   0.22    &   5.1 $\pm$   1.7 &   11  &   -61 $\pm$   10  \\
0.3 &   1.5 &   0.66    &   2.9 $\pm$   0.6 &   19  &   -74 $\pm$   6   \\
0.0   &   1.5 &   0.96    &   2.3 $\pm$   0.4 &   22  &   -87 $\pm$   5   \\
-0.3    &   1.5 &   0.82    &   2.2 $\pm$   0.5 &   18  &   79  $\pm$   6   \\
0.3 &   1.2 &   1.33    &   0.9 $\pm$   0.3 &   12  &   -71 $\pm$   9   \\
0.0   &   1.2 &   1.71    &   0.7 $\pm$   0.2 &   12  &   -83 $\pm$   9   \\
-0.6    &   1.2 &   1.23    &   1.3 $\pm$   0.3 &   16  &   19  $\pm$   7   \\
-0.9    &   1.2 &   0.56    &   3.2 $\pm$   0.7 &   18  &   -3  $\pm$   6   \\
-0.3    &   0.9 &   2.67    &   0.6 $\pm$   0.2 &   16  &   17  $\pm$   7   \\
-0.6    &   0.9 &   1.71    &   1.4 $\pm$   0.2 &   24  &   9   $\pm$   4   \\
-0.9    &   0.9 &   0.82    &   2.8 $\pm$   0.5 &   23  &   -5  $\pm$   5   \\
-1.2    &   0.9 &   0.21    &   7.5 $\pm$   1.8 &   16  &   -18 $\pm$   7   \\
0.9 &   0.6 &   0.30    &   4.7 $\pm$   1.3 &   14  &   18  $\pm$   8   \\
0.6 &   0.6 &   1.30    &   1.0   $\pm$   0.3 &   13  &   37  $\pm$   8   \\
-0.3    &   0.6 &   2.33    &   0.6 $\pm$   0.1 &   14  &   6   $\pm$   8   \\
-0.6    &   0.6 &   1.55    &   1.1 $\pm$   0.2 &   17  &   -2  $\pm$   6   \\
-0.9    &   0.6 &   0.70    &   2.7 $\pm$   0.6 &   19  &   -22 $\pm$   6   \\
0.9 &   0.3 &   0.20    &   7.5 $\pm$   1.9 &   15  &   32  $\pm$   7   \\
0.6 &   0.3 &   0.81    &   2.7 $\pm$   0.5 &   22  &   48  $\pm$   5   \\
0.3 &   0.3 &   1.75    &   0.8 $\pm$   0.2 &   14  &   65  $\pm$   8   \\
-0.9    &   0.3 &   0.32    &   3.7 $\pm$   1.1 &   12  &   -27 $\pm$   9   \\
0.6 &   0.0   &   0.22    &   8.1 $\pm$   1.7 &   18  &   55  $\pm$   6   \\
0.3 &   0.0   &   0.60    &   2.5 $\pm$   0.6 &   15  &   80  $\pm$   7   \\
0.0   &   0.0   &   0.95    &   2.1 $\pm$   0.4 &   20  &   -79 $\pm$   5   \\
-0.3    &   0.0   &   0.76    &   1.7 $\pm$   0.5 &   13  &   -80 $\pm$   9   \\
0.0   &   -0.3    &   0.23    &   6.0   $\pm$   1.6 &   14  &   -81 $\pm$   8   \\
-3.6    &   -2.4    &   0.22    &   5.0   $\pm$   1.7 &   11  &   -17 $\pm$   10  \\

\enddata
\tablecomments{\footnotesize{$\triangle$x \& $\triangle$y: offsets
in arcsecond from Right Ascension (J2000) = 19$^{\rm h}$23$^{\rm
m}$43$^{\rm s}$.95, Declination (J2000) =
14$\degr$30$\arcmin$34$\farcs$00. $I$: intensity of the Stokes $I$
component in Jy beam$^{-1}$. $P(\%)$: polarization percentage,
derived from the ratio $I_{\rm p}$/$I$. $I_{\rm p}$: the polarized
intensity in mJy beam$^{-1}$. $\phi$: position angle from the
north to the east. Listed data points are within the red box
associated with \textit{e2} in Figure 2. All listed points are
above 3 $\sigma_{I_{\rm p}}.$}}
\end{deluxetable}

\begin{deluxetable}{cccccc}

\tablecaption{SMA dust polarization at 870$\mu$m in e8}
\tablewidth{0pt} \tablehead{ \colhead{$\triangle$x} &
\colhead{$\triangle$y} & \colhead{$I$}&\colhead{$P(\%)$} &
\colhead{$I_{\rm p}$} & \colhead{$\phi(\degr)$}}

\startdata
-0.6    &   -5.1    &   0.28    &   4.0   $\pm$   1.3 &   11  &   76  $\pm$   10  \\
-1.2    &   -5.4    &   0.64    &   2.2 $\pm$   0.6 &   14  &   -1  $\pm$   8   \\
-1.5    &   -5.7    &   0.22    &   5.8 $\pm$   1.8 &   13  &   -1  $\pm$   9   \\
-0.9    &   -6.0  &   1.57    &   0.7 $\pm$   0.2 &   11  &   -70 $\pm$   10  \\
-1.5    &   -6.6    &   0.27    &   7.5 $\pm$   1.4 &   20  &   -21 $\pm$   5   \\
-1.5    &   -6.9    &   0.39    &   3.8 $\pm$   1.0   &   15  &   -19 $\pm$   7   \\
-1.2    &   -7.2    &   0.55    &   2.0   $\pm$   0.7 &   11  &   36  $\pm$   10  \\
-1.2    &   -7.5    &   0.42    &   3.3 $\pm$   0.9 &   14  &   59  $\pm$   8   \\
1.5 &   -8.7    &   0.26    &   4.2 $\pm$   1.4 &   11  &   6   $\pm$   10  \\
-1.8    &   -8.7    &   0.24    &   5.0   $\pm$   1.6 &   12  &   -33 $\pm$   9   \\
-2.1    &   -8.7    &   0.23    &   8.0   $\pm$   1.7 &   18  &   -40 $\pm$   6   \\
-2.1    &   -9.0  &   0.23    &   4.7 $\pm$   1.5 &   11  &   -33 $\pm$   10  \\

\enddata
\tablecomments{Identical notation as in Table 1. $\triangle$x and
$\triangle$y are with respect to the same coordinate as in Table
1. Listed data points are all within the red box associated with
\textit{e8} in Figure 2. }
\end{deluxetable}

\begin{deluxetable}{cc|cc|cc}

\tablecaption{Derived parameters in e2 and e8} \tablewidth{0pt}
\tablehead{\colhead{Parameters} & \colhead{units} &
\multicolumn{2}{c}{e2} & \multicolumn{2}{c}{e8}\\ & & BIMA$^{a}$ &
SMA$^{b}$ & BIMA$^{a}$ & SMA$^{b}$}

\startdata

$\theta_{maj}\times\theta_{min}$ & $\arcsec$ & 3.6 $\times$ 2.6 &
0.9 $\times$ 0.8 &
5.0 $\times$ 2.2 & 0.9 $\times$ 0.3 \\
 $M_{\rm R}$ & M$_{\sun}$ & - & 220 & - & 94 \\
 $n_{\rm H_{\rm 2}}$ & cm$^{-3}$ & 1.5 $\times$ 10$^{6}$ & 2.7 $\times$ 10$^{7}$ &
 7.6 $\times$ 10$^{5}$ & 1.2 $\times$ 10$^{7}$ \\
 $B$ & mG  &  $\geq$1 & $<$19 & $\geq$1 & $<$8 \\
 $\lambda_{\rm A}$ & mpc &  $\geq$5 & $<$6 & $\geq$11 & $<$5 \\
  & ($\arcsec$)  & ($\geq$0.2) & ($<$0.2) & ($\geq$0.3) & ($<$0.2) \\
$\lambda_{\rm T,cr}$ & mpc & 25 & 6 & 36 & 9 \\
 & ($\arcsec$) & (0.8) & (0.2) & (1.1) & (0.3) \\
$\lambda_{\rm M,cr}$ & mpc & $\geq$24 & $<$25 & $\geq$47 & $<$24 \\
& ($\arcsec$) & ($\geq$0.7) & ($<$0.8) & ($\geq$1.4) & ($<$0.7) \\
\enddata
\tablecomments{$\theta_{maj}$ and $\theta_{min}$ refer to the
major and minor axis of the deconvolved size, respectively. They
are determined by a best-fit Gaussian of the continuum emission
centered on the peak. $M_{\rm R}$ refers to the estimated $M_{\rm
gas}$ from the continuum emission within a radius of 1$\arcsec$
centered on the peak position. The lower limit of the B field
strength of 1 mG is adopted from Lai et al. (2001). Characteristic
length scales $\lambda_{\rm A}$, $\lambda_{\rm T,cr}$ and
$\lambda_{\rm M,cr}$ are calculated assuming T = 100 K. Note that
the weak constraints on $\lambda_{\rm A}$ and $\lambda_{\rm M,cr}$
result from the large possible range of B field strength.}

\vspace{2em} \footnotesize{$^{a}$The derived mean $n_{\rm H_{2}}$
is based on a Gaussian fit of the 1.3 mm continuum emission toward
the peak. All the natural length scales depend on this $n_{\rm
H_{\rm 2}}$.

$^{b}$The derived mean $n_{\rm H_{2}}$ is within a radius of 1$
\arcsec$ centered on the 870 $\mu$m continuum peak, where the
collapsing signatures were clearly revealed with the molecular
lines. All the natural length scales and the upper limit of $B$
depend on this $n_{\rm H_{\rm 2}}$.}

\end{deluxetable}

\begin{figure}
\begin{center}
\includegraphics[scale=0.7]{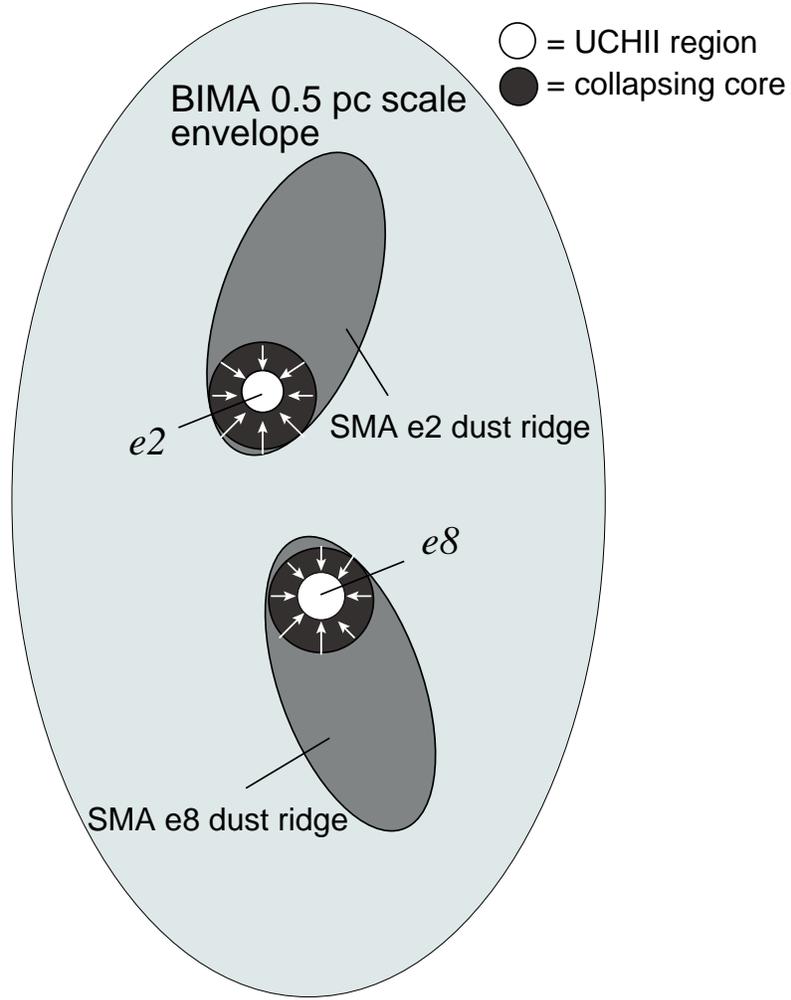}

\caption{Schematic cartoon of structures in W51 e2/e8. \textit{e2}
and \textit{e8} refer to the UCHII region e2 and e8,
respectively.} \label{schematic}
\end{center}
\end{figure}

\begin{figure}
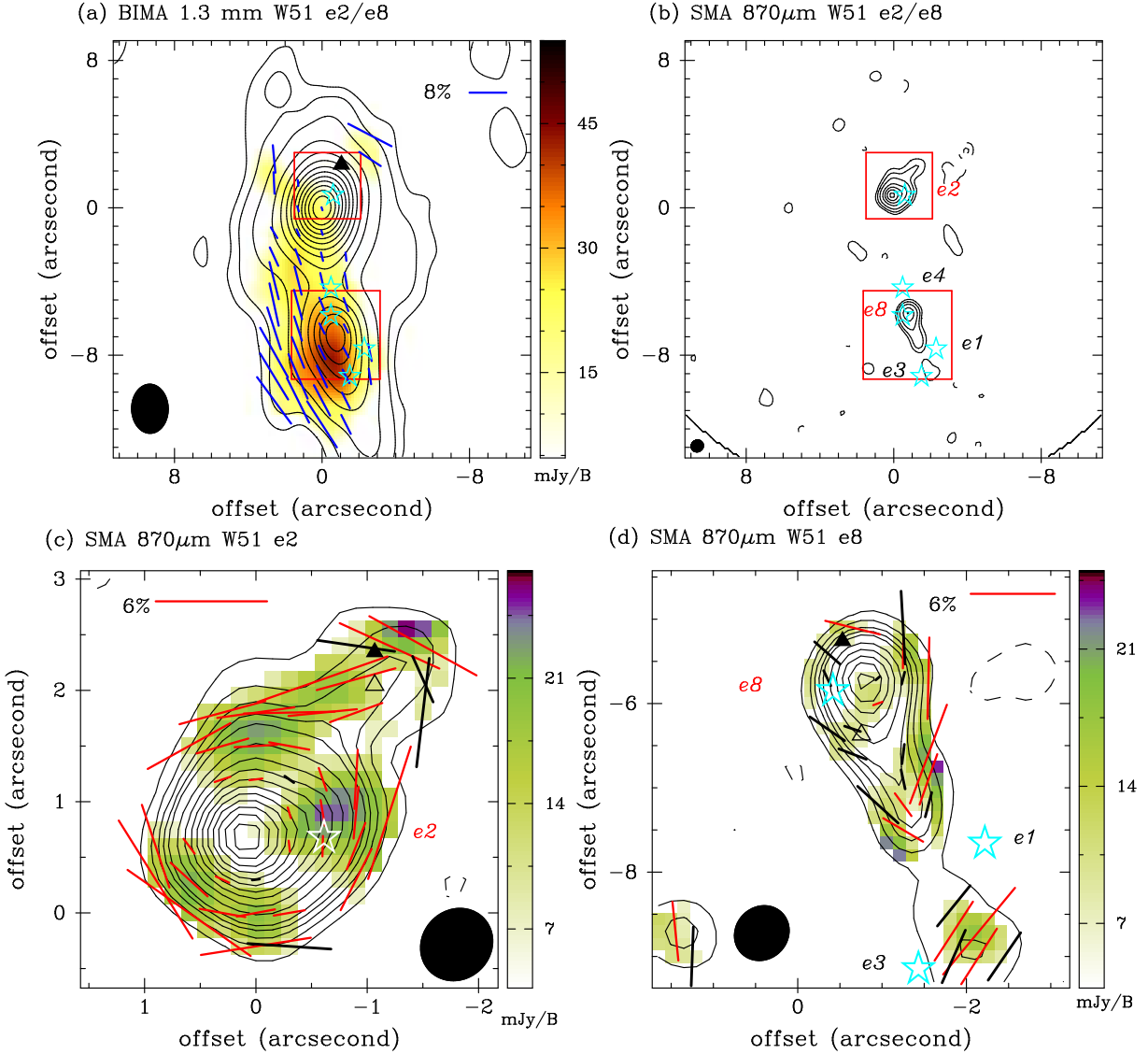

\begin{center}
\includegraphics[scale=0.6]{f2a.eps}
\includegraphics[scale=0.6]{f2b.eps}
\includegraphics[scale=0.6]{f2c.eps}
\includegraphics[scale=0.6]{f2d.eps}

\caption{\footnotesize{ (a) Polarization map observed with BIMA
adopted from Lai et al. (2001). Contours are the 1.3 mm continuum
strength of $-$3, 3, 5, 10, 20, 30, ..., 90, 100, 110 $\times$ 27
mJy beam$^{-1}$, where the size of the synthesized beam is
2$\farcs$7 $\times$ 2$\farcs$0. The cyan stars mark the UCHII
regions. The offset is in arcsecond with respect to Right
Ascension (J2000)=19$^{h}$23$^{m}$43$^{s}$.95, Declination
(J2000)=14$\degr$30$\arcmin$34$\farcs$00. Red boxes indicate the
regions presented in panel (c) and (d). (b): SMA 870 $\mu$m
continuum emission in W51 e2/e8. Contours are the 870 $\mu$m
continuum strength at $-$6, $-$3, 3, 6, 12, 24, 36, 48, 60
$\times$ 60 mJy beam$^{-1}$, where the beam size is 0$\farcs$7
$\times$ 0$\farcs$6. All the other symbols are identical to the
ones in (a). (c): SMA polarization in W51 e2. Contours are the 870
$\mu$m continuum strength at $-$4, $-$2, 2, 4, 6, 8, 10, 15, 20,
..., 55, 60 $\times$ 60 mJy beam$^{-1}$. Black and red segments
represent the polarization with its length proportional to the
polarized percentage at 2 to 3 $\sigma_{I_{\rm p}}$ and above
3$\sigma_{I_{\rm p}}$, respectively. Solid and unfilled triangles
mark the positions of the H$_{2}$O \citep{genzel81} and (J,K) =
(9,6) NH$_{3}$ \citep{pratap91} masers, respectively. All the
other symbols are identical to the ones in (a). (d) SMA
polarization map in W51 e8. All the symbols and contour levels are
identical to the ones in (c). All images shown have been corrected
for the effect of primary beam attenuation. In each panel, the
synthesized beam is plotted as a black ellipse at the lower-left
corner. In panel (a), (c) and (d), the polarized intensity is
shown in color scale with strength indicated by the wedge on the
right in units of mJy beam$^{-1}$ (mJy/B).}} \label{pol_sma_bima}
\end{center}
\end{figure}

\begin{figure}
\includegraphics[scale=1.0]{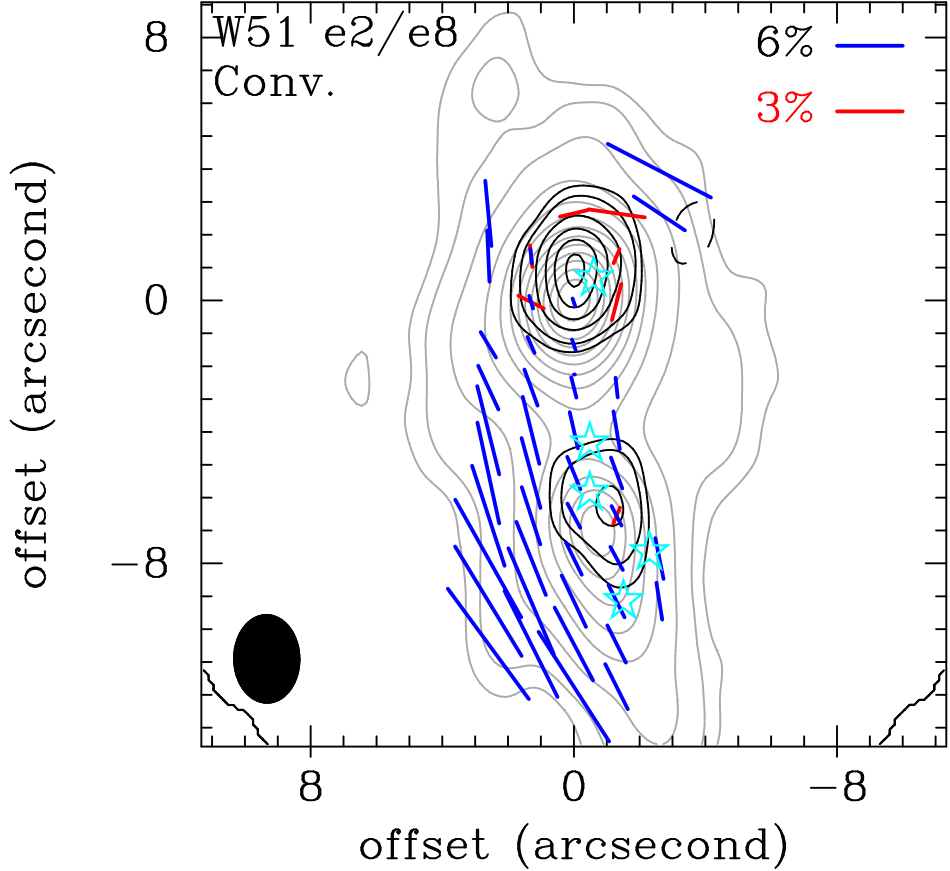}
\caption{Polarization map of the SMA results (red segments) at 870
$\mu$m restored with the BIMA synthesized beam. BIMA polarization
at 1.3 mm is shown in blue segments. The synthesized beam size is
2$\farcs$7$\times$2$\farcs$0 with a P.A. of 1$\degr$. The black
contours represent the SMA 870 $\mu$m continuum emission after
restoring to the BIMA resolution with contours of $-$3, 3, 5, 10,
15, 20, 25 $\times$ 0.3 Jy beam$^{-1}$. The grey contours
represent the BIMA 1.3 mm continuum emission with strengths as
shown in Figure 2. All polarization vectors plotted are above
3$\sigma_{I_{\rm p}}$. All the other notations are identical to
the ones in Figure 2. The slight offset between the SMA and BIMA
maps is most likely due to missing short spacing information in
the SMA data, being insensitive to structures larger than
8$\arcsec$.} \label{pol_conv}
\end{figure}

\begin{figure}
\includegraphics[scale=0.8]{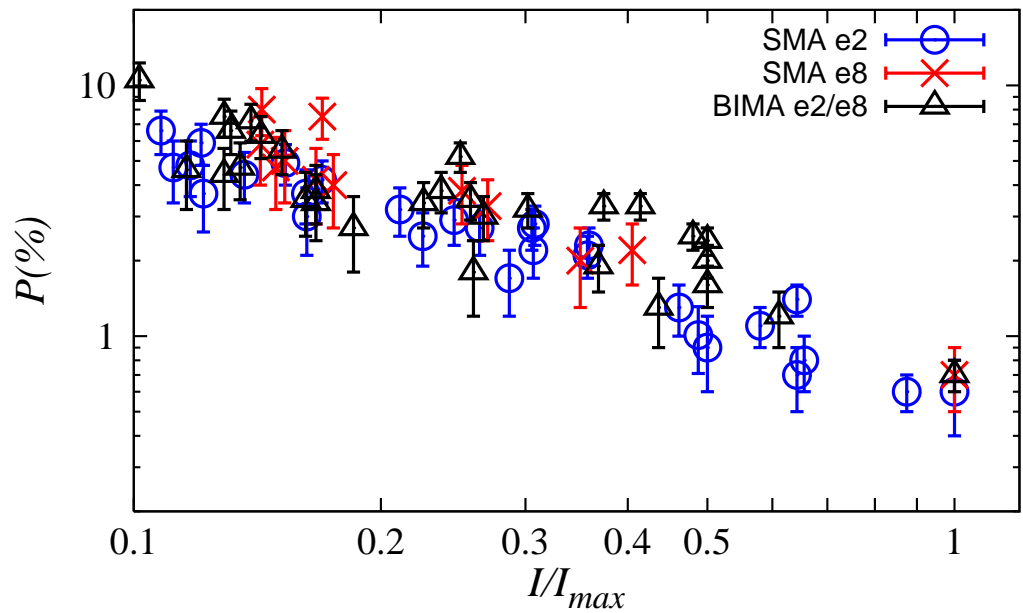}
\caption{Plot of polarization percentage, $P(\%)$, versus
normalized intensity ($I/I_{max}$). The blue circles, red crosses,
and black triangles are SMA results for e2, e8 and BIMA results
for both e2 and e8, respectively. All the plotted data are above
3$\sigma_{I_{\rm p}}$. 
}
\label{cont_conv}
\end{figure}

\newpage
\begin{figure}
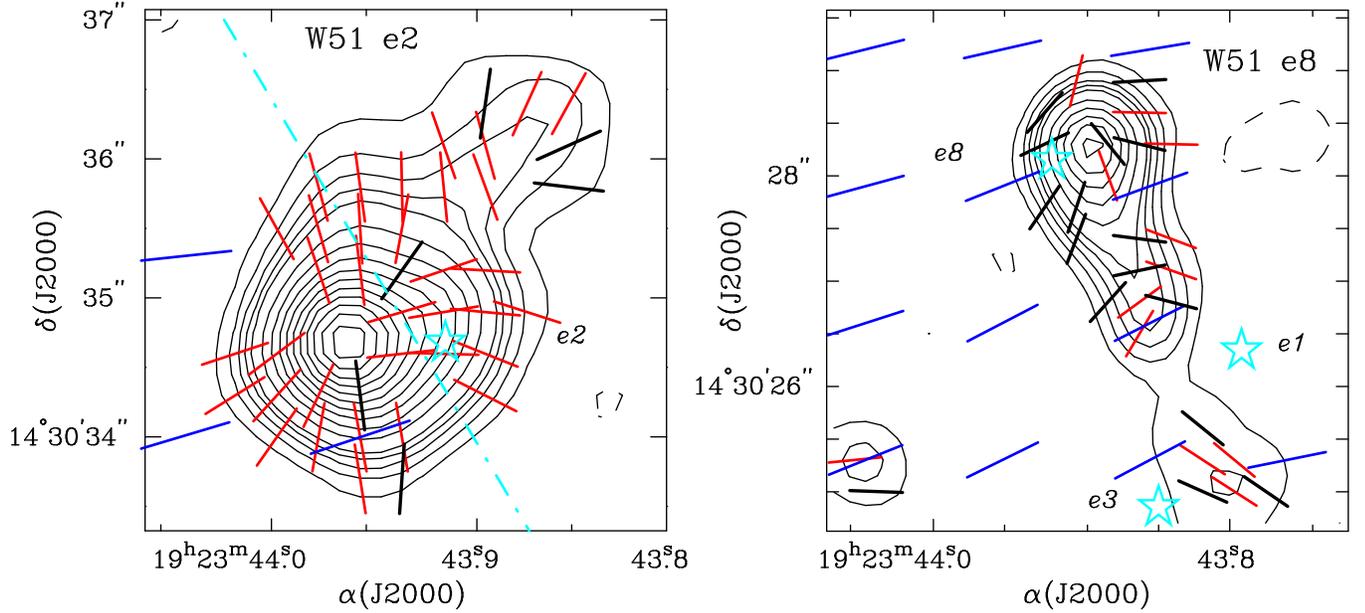

\includegraphics[scale=0.7]{f5a}
\includegraphics[scale=0.7]{f5b}

\caption{Left-panel: B field maps in W51 e2 with S/N ratio between
2 to 3 $\sigma_{I_{\rm p}}$ (black segments) and $>$3
$\sigma_{I_{\rm p}}$ (red segments), and 870 $\mu$m continuum
emission in black contours. The dot-dash cyan segment marks the
direction of the ionized accreting flow by Keto \& Klaassen
(2008). Blue segments are the inferred B field direction from the
BIMA. The magnetic field segments (all of equal length) are
rotated by 90$\degr$ with respect to the polarization segments.
Right-panel: B field map in W51 e8. All the symbols are identical
to the ones in the left-panel. All the contours are plotted with
the same level as in Figure 2(c) and 2(d) for left-panel and
right-panel, respectively. Note that even at 2 to 3
$\sigma_{I_{\rm p}}$, the B field directions are varying in a
coherent manner.  Hence the trends are significant.} \label{sma_B}
\end{figure}

\begin{figure}
\includegraphics[scale=1.2]{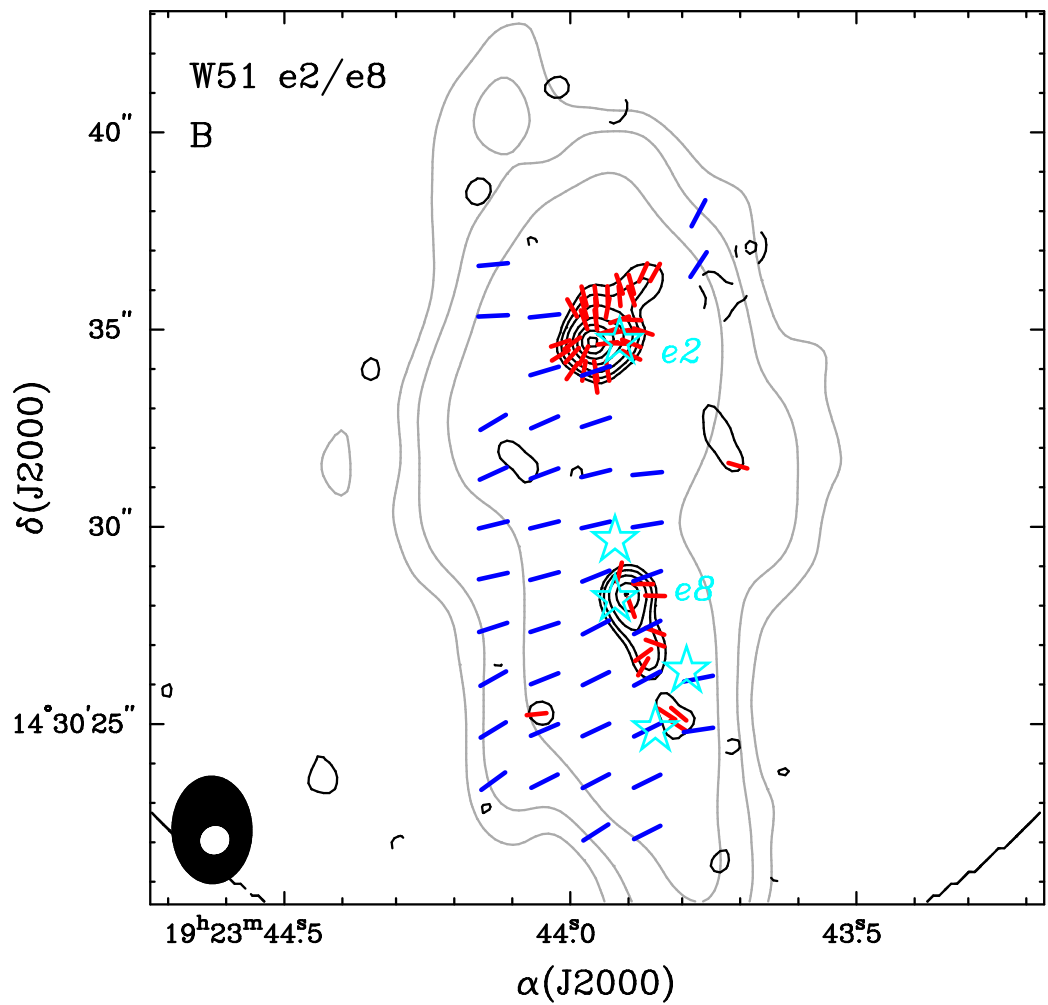}

\caption{B field maps of W51 e2/e8 from the SMA (red segments) at
870 $\mu$m and BIMA (blue segments) at 1.3 mm. The black and grey
contours represent the strengths of the continuum emission at 870
$\mu$m and at 1.3 mm, respectively. The black contours are plotted
in the same levels as in Figure 2(b), and grey contours plotted
are 3, 5, 10 $\times$ 27 mJy beam$^{-1}$. The other symbols are
identical to the ones in Figure 2. The synthesized beams of the
SMA and BIMA are plotted in the lower-left corner as white and
black ellipses, respectively.} \label{sma_bima_B}
\end{figure}

\renewcommand{\thefigure}{A\arabic{figure}}
\setcounter{figure}{0}  

\begin{figure}
\includegraphics[scale=0.6]{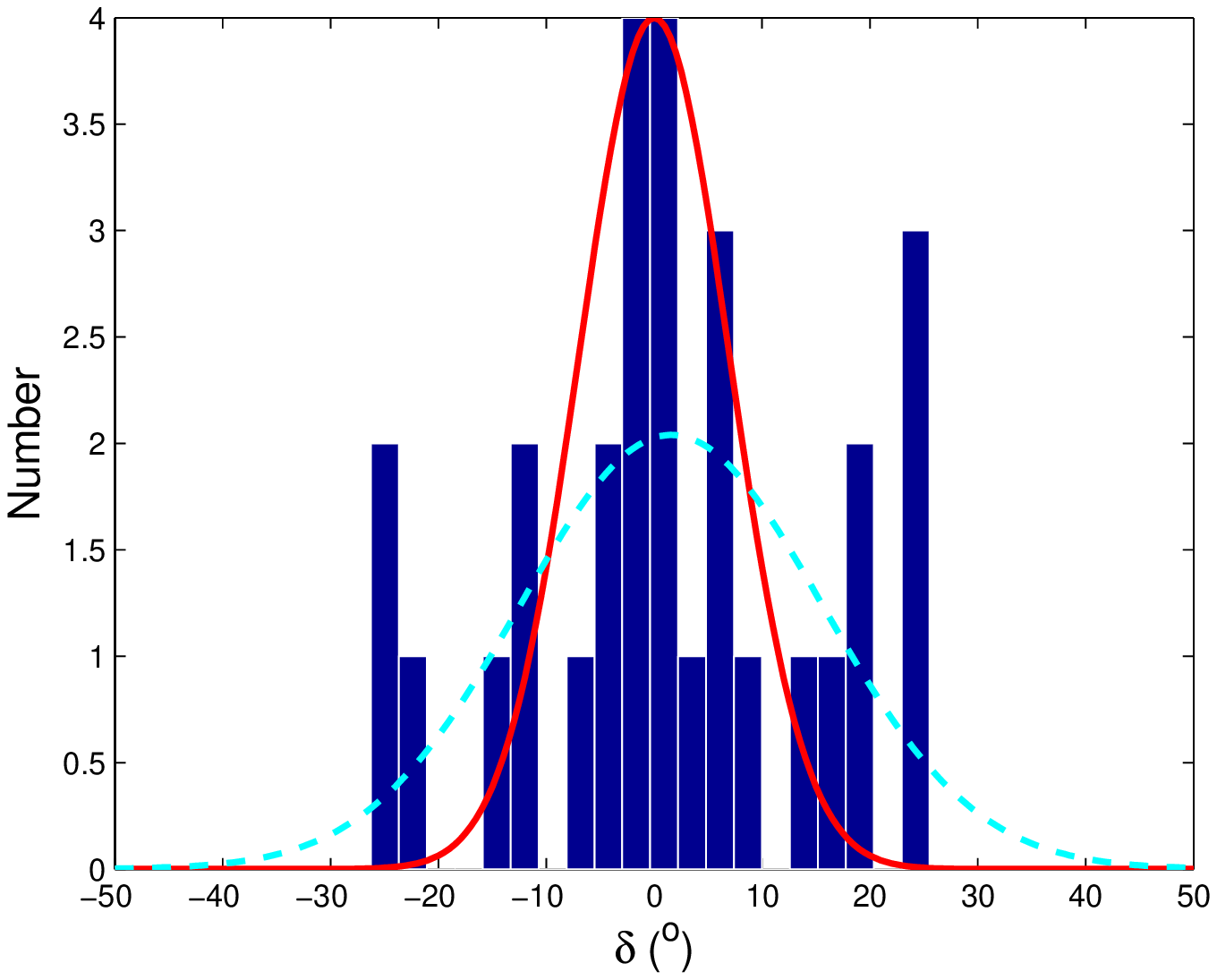}
\includegraphics[scale=0.6]{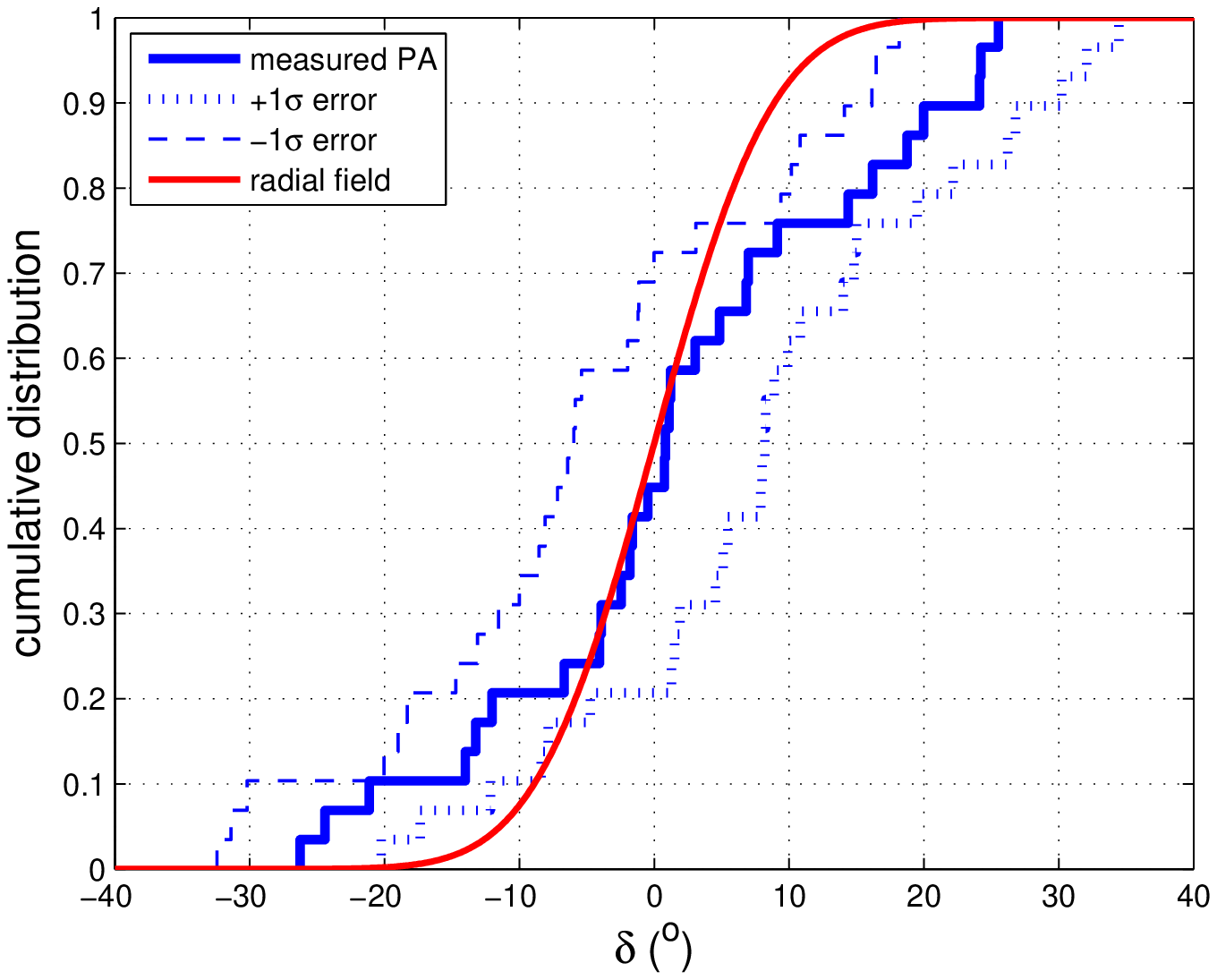}
\caption{Left-panel: Histogram of the differences $\delta$ between
the measured P.A. and a radial field. The cyan dashed curve is the
enforced fitted Gaussian to the measurement with the resultant
standard deviation $\sigma$=14$\degr$ and mean $\mu$=1.6$\degr$.
The red solid curve is the normal distribution with $\sigma$ =
$\sigma_{\rm mean, PA}$=6.9$\degr$, which is the mean uncertainty
of the measurement. Right-panel: Cumulative distribution of
$\delta$ (solid blue line). The red curve marks the cumulative
normal
distribution with $\sigma_{\rm mean, PA}$. 
The $\pm$ 1 $\sigma_{\rm mean, PA}$ error bounds from the MC
simulation, including measurement uncertainties and the shift of
the center of the radial field, are also shown.}

\label{f_kstest}
\end{figure}

\begin{figure}
\includegraphics[scale=0.8]{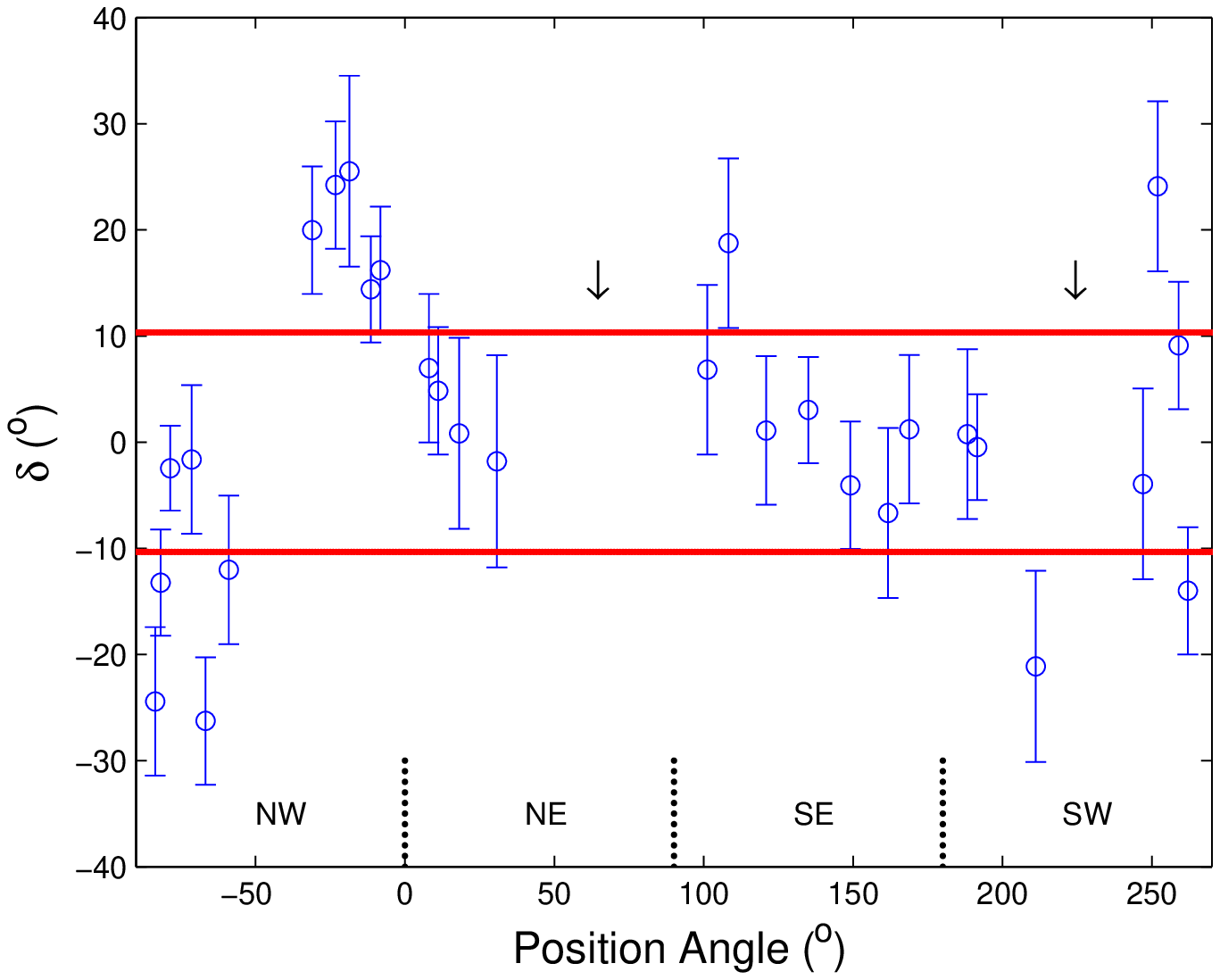}
\caption{Plot of position angle versus $\delta$. The red lines
mark the 3 $\sigma_{\rm mean, PA}$ bounds. The black arrows mark
the depolarization zones around P.A.s of 60$\degr$ and 220$\degr$,
$\sim$ parallel to the maximum velocity gradient in the
H53$\alpha$ line. The areas with respect to the 870 continuum peak
are marked as NW, NE, SE and SW, which correspond to the
north-west, north-east, south-east and south-west, respectively.
There are always $\delta$ $>$ 3 $\sigma_{\rm mean,PA}$ in the
areas separated by the depolarization zones.} \label{cont_conv}
\end{figure}

\end{document}